\documentclass[prl,aps,superscriptaddress,twocolumn,showpacs,showkeys]{revtex4-1}
\usepackage[utf8x]{inputenc}
\usepackage{microtype}
\usepackage{lmodern}
\usepackage{graphicx}
\usepackage{amsmath}
\usepackage{amsfonts}
\usepackage[colorlinks]{hyperref}

\allowdisplaybreaks[1]

\newcommand{\AF}{\lvert\mathrm{0}\rangle}
\newcommand{\FA}{\langle\mathrm{0}\rvert}

\DeclareMathOperator{\Tr}{Tr}

\begin{document}

\title{Polaron states in a CuO chain}

\author{Krzysztof Bieniasz}
\email[Corresponding author: ]{krzysztof.bieniasz@uj.edu.pl}
\affiliation{Marian Smoluchowski Institute of Physics, Jagiellonian University,
             prof. \L{}ojasiewicza 11, PL-30348 Krak\'ow, Poland }

\author{Andrzej M. Ole\'s}
\affiliation{Marian Smoluchowski Institute of Physics, Jagiellonian University,
             prof. \L{}ojasiewicza 11, PL-30348 Krak\'ow, Poland }
\affiliation{Max-Planck-Institut f\"ur Festk\"orperforschung,
Heisenbergstra{\ss}e 1, D-70569 Stuttgart, Germany}

\date{22 June, 2014}

\begin{abstract}
We introduce a one-dimensional model for a CuO chain, with holes
and $S=1/2$ spins localized in $3d_{x^2-y^2}$ orbitals, and $p_\sigma$
oxygen orbitals without holes in the ground state.
We consider a single hole doped at an oxygen site and study
its propagation by spin-flip
processes. We develop the Green's function method and treat the
hole-spin coupling in the self-consistent Born approximation,
similar to that successfully used to study polarons in the regular
$t$-$J$ model. We present an analytical solution of the problem and
investigate whether
the numerical integration is a good approximation to this solution.\\
\textit{ Published in Acta Phys. Polon. A \textbf{127}, 269 (2015).}
\end{abstract}


\maketitle

Soon after the discovery of high-$T_c$ superconductors (HTS), it was
realized that the $t$-$J$ model could be regarded as the prime candidate
for the theoretical description of the phenomenon \cite{Fuk08}.
A deeper insight into the dynamics of a hole propagation in a Mott insulator
with antiferromagnetic (AF) order might yield a clue to the understanding of the origin
of superconductivity in cuprates~\cite{Ole12}. The effective $t$-$J$ model
arises from mapping the three-band $p$-$d$ model \cite{Var87,Ole87} onto
the copper $d_{x^{2}-y^{2}}$ states and provides the simplest description of
the electronic states in CuO$_2$ planes of HTS. Recent studies indicate,
however, that oxygen $p$-orbital states strongly renormalize the
quasiparticle (QP) energy, both for antiferromagnetic \cite{Lau11} and ferromagnetic
\cite{Bie13} systems. Importance of oxygen orbitals may be recognized
both from the prominent role played in cuprates by the second neighbor
effective hopping $t'$, for instance in the stability of stripe phases
\cite{Fle01}, and from numerical studies comparing hole-doped and
electron-doped systems \cite{Ari09}.

These circumstances have led us to study the polaron dynamics in the
extended $t$-$J$ model which includes the oxygen states. In analogy to
the well known CuO$_2$ system, initially the $2p$ oxygen states are
completely filled with electrons, while the $d_{x^2-y^2}$ states are
half-filled and hosting localized $S=1/2$ spins. The superexchange $J$
couples the spins on Cu sites \cite{Zaa88} and stabilizes AF order in
the ground state. In the charge-transfer insulator there are no
itinerant charges in the $p$ states. Therefore, to activate any
kinetic energy or interaction whatsoever, one needs to inject charges
into the $p$ states, which is usually achieved by means of hole doping
La$_{2-x}$(Sr,Ca)$_x$CuO$_4$, or by substitutional transition metal
impurities of different valence than Cu ions. Here we consider the case
of a single hole injected into the $p$-band.

\begin{figure}[bt!]
  \centering
  \includegraphics[width=\columnwidth]{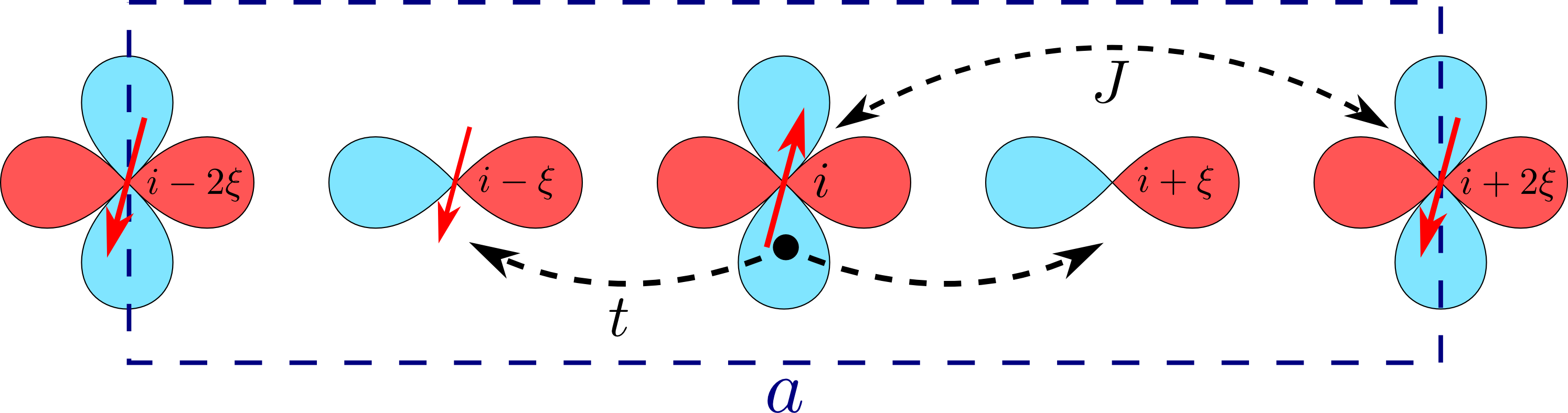}
\caption{Schematic representation of the CuO model with the conventions
used throughout this paper. The dashed line frame indicates the
magnetic unit cell in real space.}
\label{fig:model}
\end{figure}

The complexity of hopping processes in a realistic CuO$_3$ chain in
YBa$_2$Cu$_3$O$_7$ \cite{Ole91} motivates a simpler one-dimensional (1D)
model of a CuO chain, see Fig. \ref{fig:model}. The advantage of this
linear structure is that interoxygen hopping is avoided and one may
focus on the hole dynamics of a carrier injected into one of the oxygen
orbitals, which follows from three-site processes, as shown below.
Such processes are of crucial importance in systems where other hopping
processes cannot occur, as in $t_{2g}$ systems with Ising-like
superexchange, where they cause weak hole propagation \cite{Dag08} and
stabilize bond stripes \cite{Wro10}, and in the quantum compass model
\cite{Brz14}.

We formulate the simplest 1D $t$-$J$-like model with hole dynamics in
$p$ orbitals. The model Hamiltonian,
\begin{equation}
\mathcal{H} = \mathcal{H}_{0}+\mathcal{V},
\end{equation}
consists of magnetic Heisenberg exchange \eqref{eq:h0}, enforcing the
AF ground state in the ``Cu'' $d$ states and a three-site $p$-$d$-$p$
spin-flip hopping \eqref{eq:v}, obtained in second order perturbation
theory, acting as the interaction:
\begin{subequations}
  \label{eq:hamilt1}
  \begin{align}
  \label{eq:h0}
    \mathcal{H}_{0} &= J\sum_{i}
    (\mathbf{S}_{i}\cdot\mathbf{S}_{i+2\xi}-S^{2}),\\
    \label{eq:v}
    \mathcal{V} &= t\sum_{i,\sigma} (p_{i+\xi,\bar{\sigma}}^{\dag}
d_{i\bar{\sigma}}^{} d_{i\sigma}^{\dag} p_{i-\xi,\sigma}^{} + \mathrm{H.c.}),
  \end{align}
\end{subequations}
where $\xi=a/4$, $a$ is the magnetic unit cell length, the constants
$J$ (superexchange) and $t$ (hopping) are defined as positive, and the
summation extends over all the crystal unit cells. The model follows in
perturbation theory from the respective itinerant model for the CuO
chain. Note that this simple
Hamiltonian does not include any free electron kinetic terms which are
blocked due to large Coulomb repulsion on $d_{x^2-y^2}$ orbitals, and
therefore the free electrons are dispersionless. For a graphic
illustration of the model and the various conventions pertaining to it,
consult Fig.~\ref{fig:model}.

To solve this toy model in the self-consistent Born approximation
(SCBA) \cite{Mar91}, one should first notice that the pairs of
fermion operators at Cu sites are equivalent to spin-flip processes,
\begin{equation}
\label{eq:spinop}
d_{i\bar{\sigma}}^{} d_{i\sigma}^{\dag} = -S_{i}^{\sigma},
\end{equation}
where the right hand side is the spin raising/lowering operator,
depending on the value of the spin index $\sigma$. SCBA is typically
employed using the Linear Spin Wave (LSW) theory, where one expands
the spin operators in powers of bosonic operators up to the second
order using the Holstein-Primakoff representation. All the calculations
are done here in the magnetic unit cell, as the expansion of
\eqref{eq:v} around the AF ground state is irreducible to the crystal
unit cell.

After performing the standard steps of Fourier and Bogoliubov
transformations in the two sublattice framework, one arrives at the
following representation of the Hamiltonian \eqref{eq:hamilt1}:
\begin{subequations}
  \label{eq:hamilt2}
  \begin{align}
\label{eq:h2}
\mathcal{H}_{0} &= \sum_{q} \omega_{q} (\alpha_{q}^{\dag}\alpha_{q}^{}
+ \beta_{q}^{\dag}\beta_{q}^{}),\\
\label{eq:v2}
\mathcal{V} &= \frac{4t}{\sqrt{N}}\sum_{kq} \Gamma_{kq}
\left[p_{k+q,\uparrow}^{\dag} p_{k\downarrow}
(\alpha_{-q}^{\dag}+\beta_{q}) + \mathrm{H.c.}\right],
  \end{align}
\end{subequations}
where $\omega_{q} = J\lvert\sin(q/2)\rvert$ is the magnon
dispersion relation in the 1D AF model \eqref{eq:h0}, $N$ is the
number of magnetic unit cells in the system, and
\begin{equation}
\Gamma_{kq} = \sin\tfrac{k+q}{4}\sin\tfrac{k}{4} (u_{q}+v_{q})
\end{equation}
is the electron-magnon vertex function. The operators
$\{\alpha_{q},\beta_{q}\}$ are the bosonic Bogoliubov operators and
$\{u_{q},v_{q}\}$ are the Bogoliubov transformation coefficients.
More details on the above derivation can be found in the literature
concerning polarons in the $t$-$J$ model, particularly in \cite{Mar91}.

To obtain the SCBA Green's function solution, one needs to calculate
the self-energy, which then serves as the basis for the first order
Green's function:
\begin{align}
  \label{eq:self}
  \Sigma(k,\omega) &= \FA p_{k\sigma}^{} \mathcal{V}\mathcal{G}(\omega)
\mathcal{V} p_{k\sigma}^{\dag} \AF,\\
  \label{eq:resolvent}
  \mathcal{G}(\omega) &= [\omega+i\eta-\mathcal{H}]^{-1},
\end{align}
where $\AF$ is the AF ground state and $\mathcal{G}(\omega)$ is the
resolvent, or the Green's function operator. Note that, because of the
spin degrees of freedom, $\Sigma(k,\omega)$ is in fact a $2\times2$
matrix, although only in a trivial, diagonal manner. Inserting
\eqref{eq:hamilt2} into \eqref{eq:self} one quickly arrives at the
following expression for the first order self-energy:
\begin{equation}
  \label{eq:self2}
  \Sigma^{(1)}(k,\omega) = \frac{16t^{2}}{N}\sum_{q\in\mathrm{BZ}}
\frac{\Gamma_{kq}^{2}}{\omega-\omega_{q}} \mathbb{I}_{2},
\end{equation}
where $\mathbb{I}_{2}$ is a $2\times2$ identity matrix to account for
the aforementioned spin degrees of freedom and the summation extends
over the first Brillouin Zone (BZ). The above equation can be solved
analytically for the present 1D case, yielding
\begin{multline}
  \label{eq:solanal}
    \Sigma^{(1)}(k,\omega) = \tfrac{2}{\omega}\cos^{2}\tfrac{k}{4}
    \left[
      \tfrac{1}{\sqrt{1-\omega^{2}}} \ln\tfrac{\sqrt{1-\omega^{2}}-1}{\omega} +
      \ln\tfrac{2\omega-2}{\omega}
    \right]\\
     +\cos\tfrac{k}{2}
    \left[
      \tfrac{\pi}{2}
      - \tfrac{\omega}{\sqrt{1-\omega^{2}}}\ln\tfrac{\sqrt{1-\omega^{2}}-1}{\omega}
    \right],
\end{multline}
however the details of this derivation are beyond the scope of this article.

On the other hand,
the self-energy \eqref{eq:self2} can also be calculated numerically by
simple lattice summation over the BZ. After this is accomplished,
$\Sigma^{(1)}(k,\omega)$ can be used to obtain the first order Green's
function via the equation
\begin{equation}
  \label{eq:SC}
  G(k,\omega) = [(\omega+i\eta)\mathbb{I}_{2}-\Sigma(k,\omega)]^{-1},
\end{equation}
which forms the basis of the self-consistent solution. Equation
\eqref{eq:SC} can then be inserted into \eqref{eq:self} to obtain the
second order self-energy, and so on. However, usually the first order
solution suffices and further iterations do not change the result
significantly. Once the Green's function is calculated, one can extract
the physical information in the form of the spectral function
\begin{equation}
  \label{eq:spect}
  A(k,\omega) = -\frac{1}{2\pi} \Im [\Tr G(k,\omega)],
\end{equation}
which has a direct relation to photoelectron spectroscopy experiments.


\begin{figure}[bt!]
  \centering
  \includegraphics[width=\columnwidth]{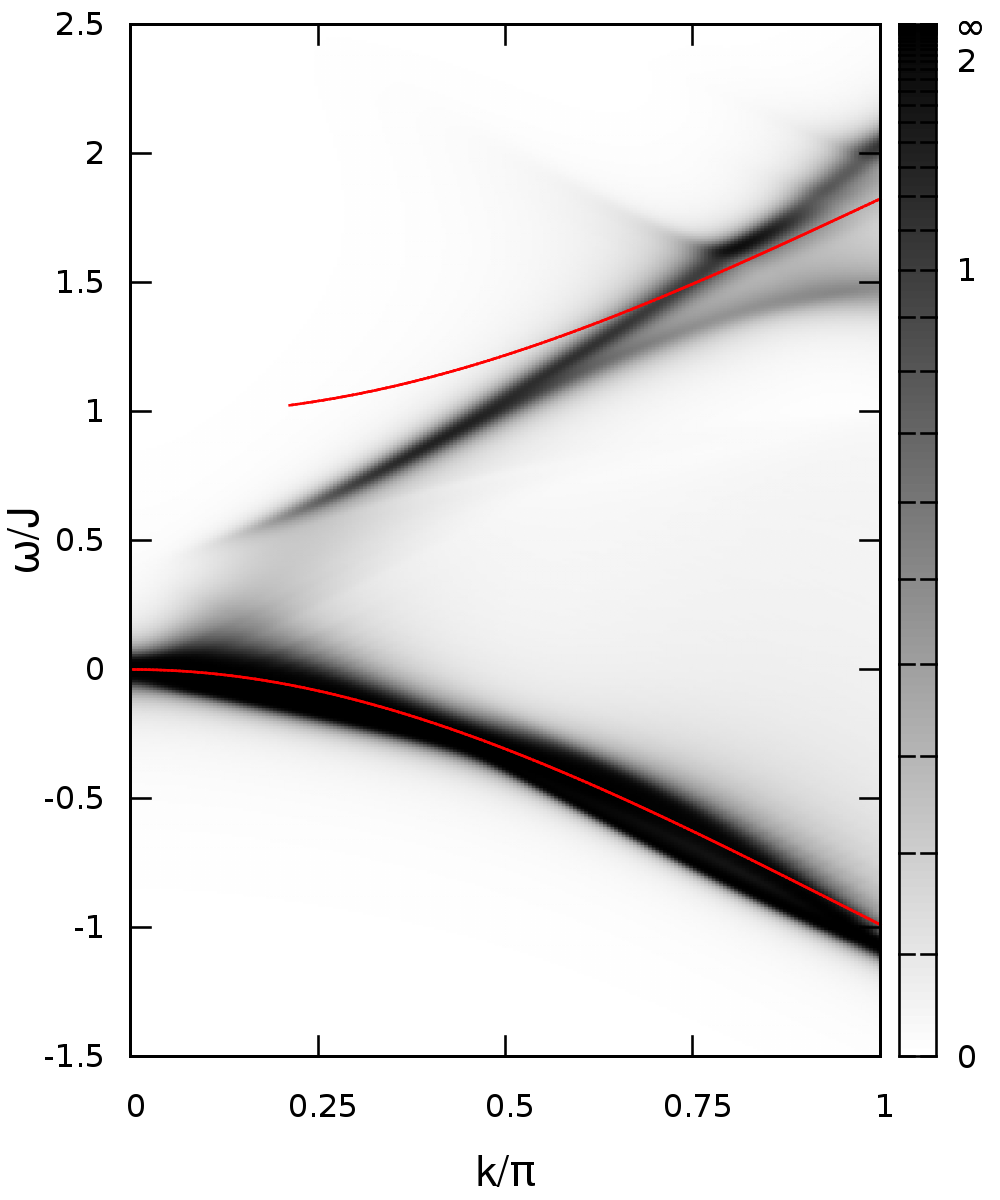}
\caption{Numerically calculated spectral function $A(k,\omega)$
(gray scale) and the analytic solution (solid red lines) for $t=1$ and
$\eta=0.01$. Note the nonlinear $\tanh$-scale.}
\label{fig:spec}
\end{figure}

Here we will present the numerical results (for self-energy in the
first order) obtained by the method described above, and contrast them
against the first order analytic result, as mentioned earlier. As can
be seen from~\eqref{eq:self2}, the natural unit of energy for this
problem is the parameter $J$, therefore the results are plotted using
the energy $\omega/J$.

Figure~\ref{fig:spec} shows the density map of the numerically obtained
spectral function $A(k,\omega)$ (in gray-scale) for the problem under
consideration with $t=1$. Bear in mind that in order to expose the
low-amplitude features of the spectrum, a nonlinear $\tanh$-scale has
been employed, hence the very strong ``broadening'' of the bands.

On top of the numerical result we have plotted the analytic solution
(solid red line). This plot was obtained by solving the equation
\begin{equation}
  \label{eq:anal}
  \omega - \Re[\Sigma^{(1)}(k,\omega+i\eta)] = 0,
\end{equation}
which defines the location of the QP bands, sans the broadening of the
spectrum, i.e., the resulting curve corresponds to the QP maxima of
\eqref{eq:spect}. Next we will discuss the differences between the two results.

First of all, one notices that both the numerical and the analytic
results display two solution branches:
($i$) the lower branch with negative energy which is a bound QP state
(since the reference free particle dispersion is 0), and
($ii$) the upper branch, which is an excited QP state --- in the present
case, as the analytic result seems to suggest, it exists only for
$k\gtrsim\pi/4$.

It is evident from Fig.~\ref{fig:spec} that in the case of the lower
branch the analytic result reproduces the numerical solution quite
well in the whole range of $k$ values. On the other hand, one can
easily notice a sharp discrepancy between the two results in the case
of the upper branch. While the analytic result ends rather abruptly
around the energy $\omega=1$, the numerical solution extends well below
this point. Since the analytic solution does not employ any
approximations (beyond the physical ones), we rather expect the
numerical approach to be unsound.

It is our supposition that the problem lies within the too na\"ive
an approach to the numerical integration. More specifically, the
function being integrated in \eqref{eq:self2} is proportional to
$(\omega-\lvert\sin(q/2)\rvert)^{-1}$, which has a singularity
line in the range $\omega\in[0,1]$. Integrating such a function
numerically using simple quadrature rules, such as the rectangle rule
in this case, introduces noticeable systematic errors into the solution.
Such errors should be especially visible in the range in which the
divergence occurs. As can be seen in Fig.~\ref{fig:spec}, this is
indeed the case since the discrepancy between the two solutions is most
pronounced exactly in the range where the singularities occur.

Further, from \eqref{eq:self2} one can see that in order to calculate
$\Sigma(k,\omega)$, one needs to integrate all the
$(k-q,\omega-\omega_{q})$ elements, which means that the errors
introduced in the lower energies propagate to the solution at higher
energies, which would explain the mismatch between the solutions also
for $\omega>1$. Moreover, this implies that a na\"ive numerical
solution of the problem presented herein is reliable only until the energy
$\omega\approx0$.

In summary, we have presented the exact analytic result and an
approximate numerical solution of the formulated problem. By comparing them we
have emphasized that an unsophisticated approach to numerical methods
can be detrimental to the reliability of the SCBA results.

\section{Acknowledgments}

We kindly acknowledge financial support by the Polish National
Science Center (NCN) under Project No.~2012/04/A/ST3/00331.

\end{document}